\documentclass[twocolumn]{aastex631}

\usepackage[table]{xcolor}
\usepackage{makecell}

\received{April 23, 2025}
\accepted{September 29, 2025}

\begin{document}


\title{Detection of OH maser emission in the 71-year periodic comet 12P/Pons–Brooks using the 40-m Thai National Radio Telescope (TNRT)}

\author[0000-0002-5814-0554]{Nobuyuki Sakai}
\affiliation{National Astronomical Research Institute of Thailand (Public Organization), 260 Moo 4, T. Donkaew, A. Maerim, Chiang Mai, 50180, Thailand}
\affiliation{Mizusawa VLBI Observatory, National Astronomical Observatory of Japan, 2-12 Hoshigaoka, Mizusawa, Oshu, Iwate 023-0861, Japan}

\author{Saran Poshyachinda}
\affiliation{National Astronomical Research Institute of Thailand (Public Organization), 260 Moo 4, T. Donkaew, A. Maerim, Chiang Mai, 50180, Thailand}


\author[0000-0002-6033-5000]{Koichiro Sugiyama}
\affiliation{National Astronomical Research Institute of Thailand (Public Organization), 260 Moo 4, T. Donkaew, A. Maerim, Chiang Mai, 50180, Thailand}

\author[0000-0002-0303-499X]{Wiphu Rujopakarn}
\affiliation{National Astronomical Research Institute of Thailand (Public Organization), 260 Moo 4, T. Donkaew, A. Maerim, Chiang Mai, 50180, Thailand}

\author[0000-0001-5511-7183]{Boonrucksar Soonthornthum}
\affiliation{National Astronomical Research Institute of Thailand (Public Organization), 260 Moo 4, T. Donkaew, A. Maerim, Chiang Mai, 50180, Thailand}

\author{Apichat Leckngam}
\affiliation{National Astronomical Research Institute of Thailand (Public Organization), 260 Moo 4, T. Donkaew, A. Maerim, Chiang Mai, 50180, Thailand}

\author[0000-0001-8168-5141]{Busaba H. Kramer}
\affiliation{Max-Planck-Institut f\"{u}r Radioastronomie, Auf dem H\"{u}gel 69, 53121 Bonn, Germany}
\affiliation{National Astronomical Research Institute of Thailand (Public Organization), 260 Moo 4, T. Donkaew, A. Maerim, Chiang Mai, 50180, Thailand}

\author{Phrudth Jaroenjittichai}
\affiliation{National Astronomical Research Institute of Thailand (Public Organization), 260 Moo 4, T. Donkaew, A. Maerim, Chiang Mai, 50180, Thailand}

\author[0000-0001-8782-0754]{Bannawit Pimpanuwat}
\affiliation{National Astronomical Research Institute of Thailand (Public Organization), 260 Moo 4, T. Donkaew, A. Maerim, Chiang Mai, 50180, Thailand}

\author{Dan Singwong}
\affiliation{National Astronomical Research Institute of Thailand (Public Organization), 260 Moo 4, T. Donkaew, A. Maerim, Chiang Mai, 50180, Thailand}

\author{Kitipoom Kanjana} 
\affiliation{National Astronomical Research Institute of Thailand (Public Organization), 260 Moo 4, T. Donkaew, A. Maerim, Chiang Mai, 50180, Thailand}

\author{Nikom Prasert}
\affiliation{National Astronomical Research Institute of Thailand (Public Organization), 260 Moo 4, T. Donkaew, A. Maerim, Chiang Mai, 50180, Thailand}

\author{Songklod Punyawarin}
\affiliation{National Astronomical Research Institute of Thailand (Public Organization), 260 Moo 4, T. Donkaew, A. Maerim, Chiang Mai, 50180, Thailand}

\author{Spiro George Sarris} 
\affiliation{National Astronomical Research Institute of Thailand (Public Organization), 260 Moo 4, T. Donkaew, A. Maerim, Chiang Mai, 50180, Thailand}

\author{Teep Chairin}
\affiliation{National Astronomical Research Institute of Thailand (Public Organization), 260 Moo 4, T. Donkaew, A. Maerim, Chiang Mai, 50180, Thailand}

\author{Kamorn Bandudej}
\affiliation{National Astronomical Research Institute of Thailand (Public Organization), 260 Moo 4, T. Donkaew, A. Maerim, Chiang Mai, 50180, Thailand}

\author{Natthaphong Kruekoch} 
\affiliation{National Astronomical Research Institute of Thailand (Public Organization), 260 Moo 4, T. Donkaew, A. Maerim, Chiang Mai, 50180, Thailand}

\author{Nattapong Duangrit} 
\affiliation{National Astronomical Research Institute of Thailand (Public Organization), 260 Moo 4, T. Donkaew, A. Maerim, Chiang Mai, 50180, Thailand}

\author{Chalunthon Nuchur}
\affiliation{National Astronomical Research Institute of Thailand (Public Organization), 260 Moo 4, T. Donkaew, A. Maerim, Chiang Mai, 50180, Thailand}

\author{Haseng Sani} 
\affiliation{National Astronomical Research Institute of Thailand (Public Organization), 260 Moo 4, T. Donkaew, A. Maerim, Chiang Mai, 50180, Thailand}

\author{Nattawit Chanwedchasart} 
\affiliation{National Astronomical Research Institute of Thailand (Public Organization), 260 Moo 4, T. Donkaew, A. Maerim, Chiang Mai, 50180, Thailand}

\author{Pathit Chatuphot}
\affiliation{National Astronomical Research Institute of Thailand (Public Organization), 260 Moo 4, T. Donkaew, A. Maerim, Chiang Mai, 50180, Thailand}

\author{Prachayapan Jiraya} 
\affiliation{National Astronomical Research Institute of Thailand (Public Organization), 260 Moo 4, T. Donkaew, A. Maerim, Chiang Mai, 50180, Thailand}

\author{Saharat Sathipjan}
\affiliation{National Astronomical Research Institute of Thailand (Public Organization), 260 Moo 4, T. Donkaew, A. Maerim, Chiang Mai, 50180, Thailand}

\author{Adirake Eakwan} 
\affiliation{National Astronomical Research Institute of Thailand (Public Organization), 260 Moo 4, T. Donkaew, A. Maerim, Chiang Mai, 50180, Thailand}

\author{Chayanin Larkaew}
\affiliation{National Astronomical Research Institute of Thailand (Public Organization), 260 Moo 4, T. Donkaew, A. Maerim, Chiang Mai, 50180, Thailand}

\author{Settasak Naewchan}
\affiliation{National Astronomical Research Institute of Thailand (Public Organization), 260 Moo 4, T. Donkaew, A. Maerim, Chiang Mai, 50180, Thailand}

\author{Pichate Pha-Kham}
\affiliation{National Astronomical Research Institute of Thailand (Public Organization), 260 Moo 4, T. Donkaew, A. Maerim, Chiang Mai, 50180, Thailand}

\author{Siritida Duangbuppha}
\affiliation{National Astronomical Research Institute of Thailand (Public Organization), 260 Moo 4, T. Donkaew, A. Maerim, Chiang Mai, 50180, Thailand}

\author{Natee PongTeerarat}
\affiliation{National Astronomical Research Institute of Thailand (Public Organization), 260 Moo 4, T. Donkaew, A. Maerim, Chiang Mai, 50180, Thailand}

\author{Nutdanai Hantankul}
\affiliation{National Astronomical Research Institute of Thailand (Public Organization), 260 Moo 4, T. Donkaew, A. Maerim, Chiang Mai, 50180, Thailand}

\author{Thodsawat Chaichana}
\affiliation{National Astronomical Research Institute of Thailand (Public Organization), 260 Moo 4, T. Donkaew, A. Maerim, Chiang Mai, 50180, Thailand}

\author{Attapon Bunwong}
\affiliation{National Astronomical Research Institute of Thailand (Public Organization), 260 Moo 4, T. Donkaew, A. Maerim, Chiang Mai, 50180, Thailand}

\author{Lalida Tantiparimongkol}
\affiliation{National Astronomical Research Institute of Thailand (Public Organization), 260 Moo 4, T. Donkaew, A. Maerim, Chiang Mai, 50180, Thailand}

\author{Wijitra Potakum}
\affiliation{National Astronomical Research Institute of Thailand (Public Organization), 260 Moo 4, T. Donkaew, A. Maerim, Chiang Mai, 50180, Thailand}

\author{Panupan Doomcome}
\affiliation{National Astronomical Research Institute of Thailand (Public Organization), 260 Moo 4, T. Donkaew, A. Maerim, Chiang Mai, 50180, Thailand}

\author{Anya Poonnawatt}
\affiliation{National Astronomical Research Institute of Thailand (Public Organization), 260 Moo 4, T. Donkaew, A. Maerim, Chiang Mai, 50180, Thailand}

\author{Nonwarit Borvornsareepirom}
\affiliation{National Astronomical Research Institute of Thailand (Public Organization), 260 Moo 4, T. Donkaew, A. Maerim, Chiang Mai, 50180, Thailand}

\author{Thanadon Paksin}
\affiliation{National Astronomical Research Institute of Thailand (Public Organization), 260 Moo 4, T. Donkaew, A. Maerim, Chiang Mai, 50180, Thailand}

\author{Warakorn Noisapung}
\affiliation{National Astronomical Research Institute of Thailand (Public Organization), 260 Moo 4, T. Donkaew, A. Maerim, Chiang Mai, 50180, Thailand}

\author{Pathorn Sathapornvajana}
\affiliation{National Astronomical Research Institute of Thailand (Public Organization), 260 Moo 4, T. Donkaew, A. Maerim, Chiang Mai, 50180, Thailand}

\author{Naphat Yawilerng}
\affiliation{National Astronomical Research Institute of Thailand (Public Organization), 260 Moo 4, T. Donkaew, A. Maerim, Chiang Mai, 50180, Thailand}

\author{Nonnadda Silamai}]
\affiliation{National Astronomical Research Institute of Thailand (Public Organization), 260 Moo 4, T. Donkaew, A. Maerim, Chiang Mai, 50180, Thailand}

\author{Ornjira Krataithong}
\affiliation{National Astronomical Research Institute of Thailand (Public Organization), 260 Moo 4, T. Donkaew, A. Maerim, Chiang Mai, 50180, Thailand}

\author{Thita Dilokthanakul}
\affiliation{National Astronomical Research Institute of Thailand (Public Organization), 260 Moo 4, T. Donkaew, A. Maerim, Chiang Mai, 50180, Thailand}

\author{Kritsada Angkaew}
\affiliation{National Astronomical Research Institute of Thailand (Public Organization), 260 Moo 4, T. Donkaew, A. Maerim, Chiang Mai, 50180, Thailand}

\author{Nattawut Chaiwongwan}
\affiliation{National Astronomical Research Institute of Thailand (Public Organization), 260 Moo 4, T. Donkaew, A. Maerim, Chiang Mai, 50180, Thailand}

\author{Sothaya Prathumsub}
\affiliation{National Astronomical Research Institute of Thailand (Public Organization), 260 Moo 4, T. Donkaew, A. Maerim, Chiang Mai, 50180, Thailand}

\author{Jittimat Khanchaiyapoom}
\affiliation{National Astronomical Research Institute of Thailand (Public Organization), 260 Moo 4, T. Donkaew, A. Maerim, Chiang Mai, 50180, Thailand}

\author[0000-0001-8066-1631]{Hideyuki Kobayashi}
\affiliation{National Astronomical Research Institute of Thailand (Public Organization), 260 Moo 4, T. Donkaew, A. Maerim, Chiang Mai, 50180, Thailand}

\author{Yu-An Chen}
\affiliation{National Astronomical Research Institute of Thailand (Public Organization), 260 Moo 4, T. Donkaew, A. Maerim, Chiang Mai, 50180, Thailand}

\author{Nawamin Intrarat}
\affiliation{National Astronomical Research Institute of Thailand (Public Organization), 260 Moo 4, T. Donkaew, A. Maerim, Chiang Mai, 50180, Thailand}

\author[0000-0002-2758-1840]{Gundolf Wieching}
\affiliation{Max-Planck-Institut f\"{u}r Radioastronomie, Auf dem H\"{u}gel 69, 53121 Bonn, Germany}


\author[0000-0001-8715-9628]{Ewan Barr}
\affiliation{Max-Planck-Institut f\"{u}r Radioastronomie, Auf dem H\"{u}gel 69, 53121 Bonn, Germany}

\author[0000-0003-3536-4368]{Jason Wu}
\affiliation{Max-Planck-Institut f\"{u}r Radioastronomie, Auf dem H\"{u}gel 69, 53121 Bonn, Germany}

\author{Niclas Esser}
\affiliation{Max-Planck-Institut f\"{u}r Radioastronomie, Auf dem H\"{u}gel 69, 53121 Bonn, Germany}

\author{Tobias Winchen}
\affiliation{Max-Planck-Institut f\"{u}r Radioastronomie, Auf dem H\"{u}gel 69, 53121 Bonn, Germany}

\author[0000-0002-5902-5005]{Pablo de Vicente}
\affiliation{IGN, Observatorio de Yebes, Cerro de la Palera S/N, E-19141, Yebes, Guadalajara, Spain}



\begin{abstract}
Although the optical cometary database is extensive, the radio database is limited. The 18-cm OH maser observations of comets allow us to determine (i) the production rate of OH ($Q_{\rm{OH}}$) and (ii) the water expansion velocity, for each comet. To reveal the physical properties of the periodic comet 12P/Pons-Brooks, we conducted the OH maser observations of the comet using the 40-m TNRT (Thai National Radio Telescope) on March 22nd, 27th and 29th, 2024 before the perihelion passage on April 21st, 2024. We successfully detected 1665 and 1667 MHz OH maser emissions from the comet. The average OH production rates of 12P/Pons-Brooks were determined as 4.28$\pm$0.30 $\times$ 10$^{29}$ sec$^{-1}$, 5.21$\pm$0.42 $\times$ 10$^{29}$ sec$^{-1}$, and 3.36$\pm$0.43 $\times$ 10$^{29}$ sec$^{-1}$ for March 22nd, 27th and 29th, respectively. Combining our results with previous estimates, we find that the OH production rate of 12P/Pons-Brooks shows some fluctuations on timescales of a few days, but gradually increases on longer timescales as the comet approaches the last perihelion. The water expansion velocities of the comet were determined to be 1.55$\pm$0.14 km s$^{-1}$, 1.55$\pm$0.35 km s$^{-1}$, and 2.02$^{+0.47}_{-0.45}$ km s$^{-1}$ for the observations on March 22nd, 27th and 29th, respectively. The reason why the errors are different is because the signal to noise ratio is different for each OH maser emission. All expansion velocities are consistent with the individual results of the heuristic law \citep{2007AA...467..729T} within errors.



\end{abstract}









\section{Introduction} \label{sec:intro}






Comets retain the information about the early stages of the solar system formation. While more than 20 chemical species have been detected in bright comets, the main volatile constituent of comets is water (e.g., see \citealp{2011ARA&A..49..471M, 2024come.book..459B}). OH radicals are thought to be created by the photodissociation of water when comets approach the Sun. OH molecules in the cometary coma (i.e., atmosphere) radiates at ultraviolet, near infrared, 18-cm (radio) and submillimeter bands (e.g., see \citealp{1988ApJ...331.1058S}). After the first discovery of 18-cm cometary OH maser in Comet Kohoutek (1973f) (\citealt{1974A&A....34..163B}; \citealt{1974ApJ...189L.137T}), 18-cm OH masers have been detected in more than 65 comets \citep{2002IAUS..206..443C}. OH molecules absorb solar ultraviolet photons and cascade back to the ground-state $\Lambda$-doublet of OH by subsequent fluorescence \citep{1974A&A....34..163B}. The relative population inversion of upper and lower energy levels of the OH ground state depends on the heliocentric velocity (i.e., the Swings effect; \citealp{1974ApJ...191L.145M}). Typical cometary OH maser flux density ranges between 0.01 and 0.05 Jy, although a bright cometary OH maser can reach 0.4 Jy or more \citep{2022A&AT...33..139S, 2002A&A...393.1053C}. Note that the cometary OH maser flux density depends on the cometary water production rate and the observation conditions (i.e., topocentric distance and maser inversion; see \citealp{1990A&A...238..382B}). Cometary OH masers can be detected at 1612, 1665, 1667 and 1720 MHz. The statistical weight ratio of 1612, 1665, 1667 and 1667-MHz emissions is 1:5:9:1 (e.g., \citealp{1985A&A...146....1G}). The integrated intensity of the 18-cm OH line allows us to determine the OH production rate ($Q_{\rm{OH}}$) of each comet (e.g., see \citealp{2002A&A...393.1053C}). Since OH is a photo-dissociation product of H$_{2}$O, water production rate ($Q_{\rm{H_{2}O}}$) is proportional to $Q_{\rm{OH}}$ \citep{1989A&A...213..459C}. The water production rate is an important value in cometary studies because water is typically the most abundant parent gas species in cometary comas, and measurements of all other species are compared to water for compositional interpretation \citep{2009AJ....137.4734C}. A molecular production rate can be used to examine the comet's outgassing activity. The 18-cm OH line profile of each comet can be used to measure the water expansion velocity in the comet's coma \citep{1990A&A...238..382B}. The expansion velocity information can allow us to investigate the kinematics of comet's atmosphere.



\begin{figure}[tbhp] 
 \begin{center} 
     \includegraphics[scale=0.23]{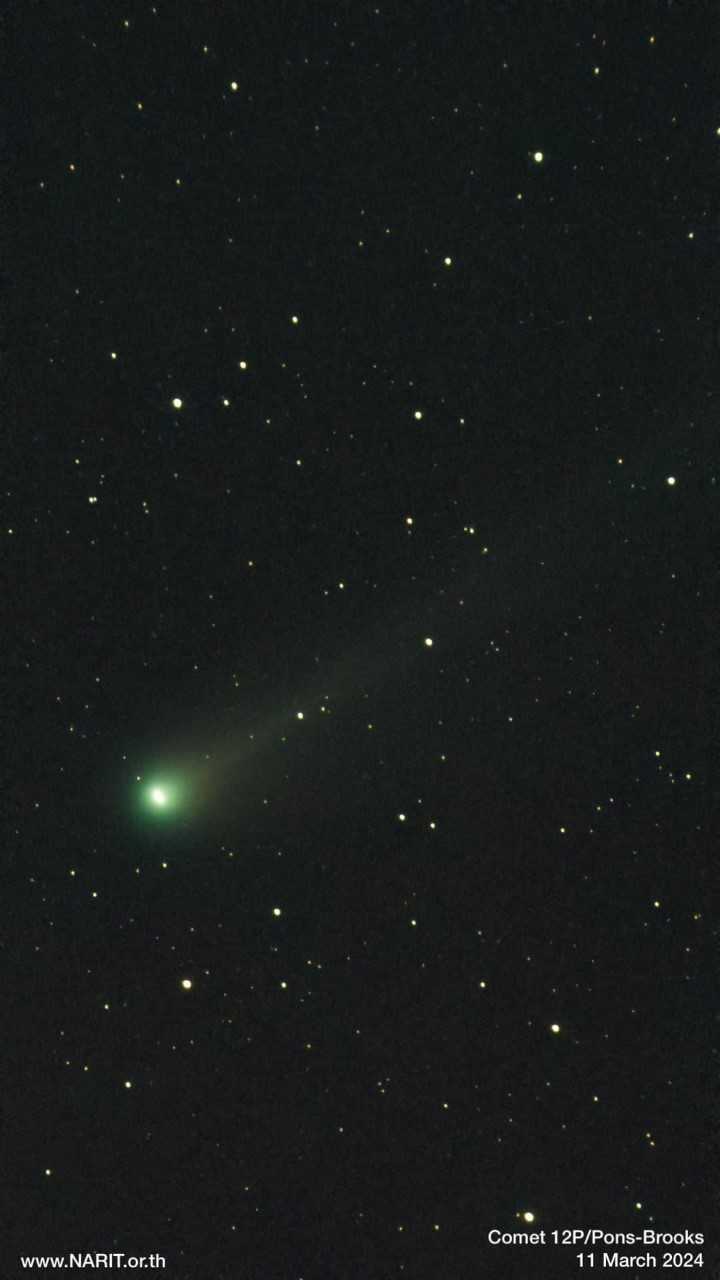} 
\end{center} 
\caption{Comet 12P/Pons–Brooks as seen from Doi Inthanon National Park, Chiangmai, Thailand on March 11th, 2024. A 9-minute exposure image was taken using a ZWO See Star S50 camera with a 50 mm f/5 lens held on an Alt-azimuth tracking mount. The image is approximately 3.7 deg $\times$ 2.1 deg in total. Image credit: NARIT Outreach Team.}
\label{fig:0} 
\end{figure}

Comet 12P/Pons-Brooks was discovered by astronomer Jean-Louis Pons in 1812 and rediscovered by astronomer William Robert Brooks in 1883 \citep{1986AJ.....91..971Y}. 12P/Pons-Brooks has an orbital period of 71.24 years\footnote{Based on the JPL Horizons System:\url{https://ssd.jpl.nasa.gov/horizons/}\label{fot:NASA}}, and thus it is classified as a Halley-type comet. The comet approached the last perihelion on April 21, 2024 (UTC). Figure \ref{fig:0} represents an optical image of comet 12P/Pons-Brooks taken on 
March 11th, 2024. It shows typical cometary characteristics such as the coma and a tail.
 Optical outbursts (i.e., sudden and intense surges in the optical brightness) of the comet have been reported during the comet appearances in 1883, 1954, and 2024 (\citealp{1883AN....107..131C,1955MNRAS.115..190P,2023ATel16194....1M,2024ATel16498....1J}). The outburst events involve the rapid release of large amounts of gas and dust from the comet's nucleus \citep{2025MNRAS.tmp..295G}. The ejected mass from 12P/Pons-Brooks is thought to be responsible for the weak December $\kappa$-Draconids meteor
shower, which typically occurs from November 29th to December 13th \citep{2016A&A...592A.107T}.



\begin{table*}[htbp] 
\begin{center}
\caption{TNRT observation information of 12P/Pons–Brooks} 
\label{table:1}
\small 
\begin{tabular}{llcccccc} 
\hline 
\hline 
R.A. (J2000)\tablenotemark{$*$}&Decl. (J2000)\tablenotemark{$*$} &Time range&$\Delta\tablenotemark{$*$}$ &$r_{h}\tablenotemark{$*$}$	&$v_{\Delta}\tablenotemark{$*$}$&$v_{h}\tablenotemark{$*$}$ &Coma diameter\tablenotemark{$\dag$} \\
	       
hh:mm:ss &dd:':" &(UTC)&(au)&(au)&(km s$^{-1}$)&(km s$^{-1}$)&(arcminute)\\ 
\hline
\multicolumn{2}{l}{March 22nd, 2024 (UTC)} &&&&&&7.5$\pm$3.4\\
01:24:33.6  	&+28:18:21	  & 05:06 - 05:34&1.620&0.960&$-$2.7&$-$18.1\\
01:24:39.5  	&+28:17:44	  & 05:43 - 06:00 &1.620&0.960&$-$2.7&$-$18.1\\
01:24:45.5  	&+28:17:08	  & 06:02 - 06:29 &1.620&0.959&$-$2.6&$-$18.1\\
01:24:51.4  	&+28:16:32	  & 06:32 - 07:00 &1.620&0.959&$-$2.6&$-$18.1\\
01:24:57.4  	&+28:15:56	  & 07:04 - 07:31 &1.620&0.959&$-$2.5&$-$18.1\\
01:25:03.3  	&+28:15:20	  & 07:33 - 08:01 &1.620&0.958&$-$2.5&$-$18.0\\
01:25:09.3  	&+28:14:44	  & 08:03 - 08:30 &1.620&0.958&$-$2.4&$-$18.0\\
01:25:15.2  	&+28:14:07	  & 08:33 - 09:01 &1.620&0.958&$-$2.4&$-$18.0\\
\multicolumn{2}{l}{March 27th, 2024 (UTC)}&&&&&&6.5$\pm$2.4 \\
01:47:46.3  	&+25:46:04	  & 04:32 - 04:59 &1.614&0.910 &$-$1.7&$-$16.2\\
01:47:52.0  	&+25:45:23	 &  05:00 - 05:30 &1.614&0.910&$-$1.6&$-$16.2 \\
01:47:57.7  	&+25:44:43	 &  05:31 - 05:59 &1.614&0.910&$-$1.6&$-$16.2 \\
01:48:03.4  	&+25:44:03	 &  06:03 - 06:31 &1.614&0.910&$-$1.5&$-$16.2 \\
01:48:09.0  	&+25:43:22	 &  06:33 - 07:01 &1.614&0.909&$-$1.5&$-$16.2 \\
01:48:14.7  	&+25:42:42	 &  07:03 - 07:29 &1.614&0.909 &$-$1.4&$-$16.2\\
01:48:20.4  	&+25:42:02	 &  07:31 - 08:03 &1.614&0.909 &$-$1.4&$-$16.2\\
01:48:26.1  	&+25:41:21	 &  08:07 - 08:30 &1.614&0.909 &$-$1.3&$-$16.2\\
01:48:31.8  	&+25:40:41	 &  08:34 - 08:59 &1.614&0.909 &$-$1.3&$-$16.2\\
01:48:37.4  	&+25:40:00	 &  09:01 - 09:27 &1.614&0.908 &$-$1.2&$-$16.1\\

\multicolumn{2}{l}{March 29th, 2024 (UTC)}&&&&&&7.2$\pm$2.0 \\
01:56:48.4 	&+24:40:19	 & 04:35 - 04:58&1.613&0.892 &$-$1.3&$-$15.4\\
01:56:53.9 	&+24:39:38	 & 05:00 - 05:28&1.613&0.892&$-$1.3&$-$15.3 \\
01:56:59.5 	&+24:38:56	 & 05:30 - 05:58&1.613&0.891&$-$1.2&$-$15.3 \\
01:57:05.1 	&+24:38:14	 & 06:00 - 06:31&1.613&0.891&$-$1.2&$-$15.3 \\
01:57:10.7 	&+24:37:32	 & 06:33 - 06:57&1.613&0.891&$-$1.1&$-$15.3 \\
01:57:16.3 	&+24:36:50	 & 07:05 - 07:26&1.613&0.891&$-$1.1&$-$15.3 \\
01:57:21.8 	&+24:36:08	 & 07:34 - 08:00&1.613&0.891&$-$1.0&$-$15.3 \\
01:57:27.4 	&+24:35:26	 & 08:02 - 08:29&1.613&0.891 &$-$1.0&$-$15.3\\
01:57:33.0 	&+24:34:44	 & 08:30 - 08:58&1.613&0.890 &$-$0.9&$-$15.3\\
01:57:38.6 	&+24:34:02	 & 09:00 - 09:32&1.613&0.890 &$-$0.9&$-$15.3\\
\hline 
\multicolumn{8}{@{}l@{}}{\hbox to 0pt{\parbox{150mm}{\normalsize
\par\noindent
Columns 1-2: Right ascension (column 1) and declination (column 2) (J2000) of 12P/Pons–Brooks; Column 3: Observation time range (UTC); Columns 4-5: Topocentric and heliocentric distances of 12P/Pons-Brooks, respectively; Columns 6-7: Predicted topocentric and heliocentric velocities of 12P/Pons-Brooks, respectively; Column 8: Mean coma diameter and its standard deviation.
\tablenotetext{$*$}{These values refer to the JPL Horizons System \citep{1996DPS....28.2504G}.}
\tablenotetext{$\dag$}{Each value is the average of optical observations from the COBS International Database (\url{https://www.cobs.si}).}
}\hss}}
\end{tabular} 
\end{center} 
\end{table*}

Although intensive optical observations of 12P/Pons-Brooks have been conducted, radio observations of the comet are limited. In this paper, we report the OH maser detection in 12P/Pons-Brooks. The comet observations with the 40-m TNRT (Thai National Radio Telescope) are summarized in section 2. The TNRT data reduction is explained in section 3.
The observation results with TNRT are shown and discussed in section 4. Especially, we discuss the physical parameters of the comet, which are (i) the OH production rate and (ii) the H$_{2}$O expansion velocity. We summarize this paper in section 5.




\section{Observations} \label{sec:obs}


\begin{figure*}[tbhp] 
 \begin{center} 
     \includegraphics[scale=1.2]{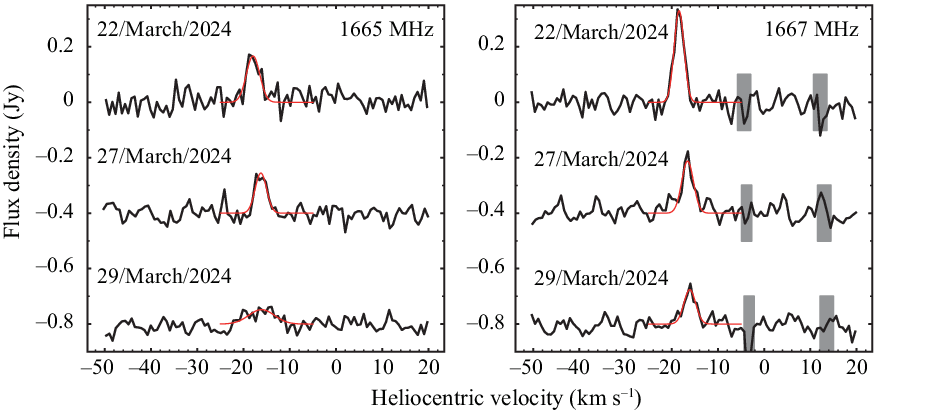} 
\end{center} 
\caption{Main-line OH maser emissions from comet 12P/Pons–Brooks. Left panel shows OH maser spectra at a rest frequency of 1665.4018 MHz, which were taken with 40-m TNRT on March 22nd (Top), 27th (Middle), and 29th (Bottom), 2024. For visualization, the middle and bottom spectra are plotted with an offset from the top spectrum. Horizontal axis shows heliocentric velocity. Red curves, showing Gaussian fits to the observed spectra (see Table \ref{table:2}), are superimposed on the spectra. Right panel is the same as the Left, but for OH maser spectra at a rest frequency of 1667.359 MHz. Shaded regions indicate radio frequency interference (RFI) at the 40-m TNRT site. Thanks to the averaging of dual polarization data, the effects of RFI are reduced.}
\label{fig:1} 
\end{figure*}

During Director’s Discretionary Time, we observed comet 12P/Pons-Brooks on March 22nd, 27th, and 29th, 2024 (UTC) with the L-band\footnote{The TNRT L-band receiver was developed in collaboration with MPIfR.}\label{fot:TNRT_Lband} (1$-$1.8 GHz; dual linear polarization) spectrometer mode of 40-m TNRT (see Table \ref{table:1}). The details of the 40-m TNRT and the L-band spectroscopy mode are summarized on the TNRT web page\footnote{\url{https://indico.narit.or.th/event/218/}\label{fot:TNRT}}. The 40-m TNRT is located at $99\fdg2170$ E, $18\fdg8643$ N and 411 m above sea level. The coordinates of the comet were updated every 30 (min) based on the ephemeris\footnote{Taken from The Sky Live:\url{https://theskylive.com/}} provided by the JPL Horizons System\footref{fot:NASA}, and the updated coordinates were applied to the observations. This is because, although TNRT can track objects based 
on equatorial coordinates, a tracking system for solar system objects based on orbital 
elements is currently under evaluation. On average, a coordinate shift was 87 (arcsec) every 30 (min), which is $\sim$7$\%$ of the TNRT beam size at 1.6 GHz ($\theta$ $\sim$ 1,200 arcsec). Considering the coordinate shift and (Azimuth, Elevation) pointing accuracy of the telescope ($\Delta$Az = 32 arcsec, $\Delta$El = 48 arcsec), for most of each observation, the comet is expected to be observed in the inner 10$\%$ of the beam. This results in $\sim$3$\%$ flux loss in a Gaussian beam, which is less than the typical amplitude calibration accuracy for TNRT L-band data (i.e., $\sim$10$\%$). The amplitude calibration accuracy of TNRT L-band data is summarized on the TNRT web page\footref{fot:TNRT}, where the brightness temperature of the H I standard region S8 \citep{1973A&AS....8..505W} obtained with TNRT is compared with brightness temperature values of the same source obtained with Effelsberg 100m \citep{2016A&A...585A..41W}, Duingeroo 25m \citep{2005A&A...440..775K}, and Parkes 64m \citep{2015A&A...578A..78K}, respectively.

TNRT front-end and back-end systems were developed by Max-Planck-Institut f\"{u}r Radioastronomie (MPIfR) and summarized on the TNRT web page\footref{fot:TNRT}. Radio frequency (RF) between 1.0 and 1.8 GHz is received by the cryogenic TNRT L-band receiver. The received RF is divided into horizontal and vertical polarization signals with an orthomode transducer (OMT). Each polarization signal is directly digitized with 8-bit sampling in the receiver box. The digitized signal is sent to the TNRT Universal Software Backend (USB) system using the Effelsberg Direct Digitization (EDD)\footnote{\url{https://mpifr-bdg.pages.mpcdf.de/edd_documentation/}\label{fot:EDD}}, where the signal is packetized and transmitted to the spectrometer via a 40 Gigabit Ethernet (GbE) Local Area Network (LAN). The output of the spectrometer is recorded on a data storage with the raw data format of Multi-Beam FITS (MBFITS; \citealp{2006A&A...454L..25M}).  

In each MBFITS file, four types of output were recorded with combinations of polarization (Vertical or Horizontal) and noise-source injection (ON or OFF). The MBFITS file consists of subscans for which we applied each integration time of 1.073 (sec), which results in a total integration time of $\sim$1 (min) for each MBFITS file. The noise source was injected for each subscan with a fraction 0.25 $\%$ of the subscan integration time (i.e., 1.073 sec $\times$ 0.25 $\%$ $\sim$ 0.003 sec).

Regarding the setup of the three observations, a frequency resolution of 3.815 kHz was applied, which corresponds to a velocity resolution of 0.686 km s$^{-1}$ at 1667.359 MHz (i.e., at the OH ground state line $^{2}\Pi_{3/2}$ (J = 3/2) of F = 2-2 ). The effective on-source times at (1665, 1667) MHz were (41, 52), (151, 179) and (143, 168) (min) on March 22nd, 27th, and 29th, 2024, respectively. These times are shorter than the total observation times (see Table \ref{table:1}) due to (1) sky (off-source) observations, (2) downtime from system processes such as output file creation and antenna movement, (3) radio frequency interference (RFI), and (4) data recording issues, which led to 50$\%$ sky data loss in the first epoch. Before and after each observation, the sky data were acquired at (Az, El) = (60, 60) and (300, 60) degrees so that the total integration time of the sky data was equal to that of the target data. Different on-source times at (1665, 1667) MHz in each day are due to RFI.

\section{Data reduction} \label{sec:analysis}
The MBFITS files obtained from the observations were analyzed using Python libraries (e.g., NumPy (\citealp{harris2020array}); Astropy (\citealp{2013A&A...558A..33A,2018AJ....156..123A})). We applied the standard analysis procedures for centimeter spectral line data to the TNRT data (e.g., see \citealp{2005ApJS..158..178M}) as follows:

\begin{enumerate}
  \item Target / Sky: $\rho$ = ($P_{\rm{target}}$/$P_{\rm{sky}}$)$-$1 
  \item The determination of opacity corrected system noise temperature: \par $T_{\rm{sys}}^{*}$ [K] = $\frac{T_{\rm{NS}}}{\left(\frac{P_{\rm{target+NS}}}{P_{\rm{target}}}-1\right)}$exp($\frac{\tau_{\rm{0}}}{\rm{sinEl}}$) 
  \item The determination of opacity corrected antenna temperature: $T_{\rm{a}}^{*}$ [K] = $\rho$ $\times$ $T_{\rm{sys}}^{*}$
  \item Unit conversion from Kelvin to Jansky: \par $F_{\nu}$ [Jy] = $\frac{2k}{A_{\rm{e}}\rm{(El)}}$ $T_{\rm{a}}^{*}$ $\times$ 10$^{26}$
  \item The Doppler effect correction
  \item The average of all the calibrated files (including dual-polarization data)
\end{enumerate}
In step 1, the radio power of target data is divided by that of sky data for the calibration of the noise floor (i.e., sky subtraction). To make the Target/Sky data flat, baseline subtractions were performed with a 3rd order polynomial for individual Target/Sky data. The value of $\rho$ should be close to zero in the absence of astronomical signals.

In step 2, $T_{\rm{NS}}$, $P_{\rm{target+NS}}$, $P_{\rm{target}}$, $\tau_{\rm{0}}$, and El are the temperature of the noise source equipped in the TNRT L-band receiver, radio power received by the target and noise source, radio power received by only the target, the opacity at the zenith, and the target elevation, respectively. Based on the TNRT web page\footref{fot:TNRT}, we fixed $T_{\rm{NS}}$ and $\tau_{\rm{0}}$ to 31.9 (K) and 0.0106, respectively. In step 3, we can determine the opacity corrected antenna temperature $T_{\rm{a}}^{*}$ (K) by multiplying $\rho$ (step 1) by $T_{\rm{sys}}^{*}$ (step 2). In step 4, $T_{\rm{a}}^{*}$ containing the target signal is shown in the unit of Jansky (Jy) where $k$ and $A_{\rm{e}}\rm{(El)}$ are the Boltzman constant and effective aperture area as a function of elevation, respectively. We applied a cubic function provided on the TNRT web page\footref{fot:TNRT} to $A_{\rm{e}}\rm{(El)}$.

In step 5, the Doppler effect during the observations was corrected\footnote{To display the TNRT data in a heliocentric velocity frame, we added the ($v_{h}$ $-$ $v_{\Delta}$) values to the Radio Frequency at the TNRT site.}\label{fot:JPL-Horizons-System} by referring to $v_{\Delta}$ and $v_{h}$ values in Table \ref{table:1}. To calculate $v_{\Delta}$ and $v_{h}$ values in the JPL Horizons System\footref{fot:NASA}, we entered (1) TNRT station coordinates (latitude, longitude, altitude), (2) target coordinates in J2000, and (3) observation time. We integrated (averaged) the calibrated files of dual-polarization data to detect a weak emission in step 6. Since most RFIs were polarized, the effect of RFIs was reduced by averaging the dual-polarization data in the last step.  






\begin{table*}[htbp] 
\begin{center}
\caption{Gaussian fits to main-line OH maser emissions from 12P/Pons–Brooks} 
\label{table:2}
\small 
\begin{tabular}{ccccccc} 
\hline 
\hline 
Date & $\nu$ & $F_{\nu}$ & $\sigma$ 	  & $v_{h}$ & FWHM & Line ratio  \\
	 (mmdd)      &(MHz) &(mJy) &(mJy)&(km s$^{-1}$)&(km s$^{-1}$)& \\ 
\hline
0322	&1665  	&167$\pm$24	&32	&$-$17.9$\pm$0.2 &3.4   & $-$\\
0322&1667&331$\pm$21	&33 &$-$18.4$\pm$0.1			&2.5 &0.69$\pm$0.10\\
\hline
0327	&1665  	&145$\pm$19	&25		&$-$16.3$\pm$0.2 &2.9 & $-$\\
0327&1667&188$\pm$21	&28	&$-$16.6$\pm$0.2&3.0   &0.62$\pm$0.11\\
\hline
0329	&1665  	&52$\pm$10	&21			&$-$16.1$\pm$0.6 &6.5  &$-$\tablenotemark{$*$}  \\
0329&1667&124$\pm$14	&23	&$-$16.1$\pm$0.2&3.0  & $-$\tablenotemark{$*$} \\
\hline 
\multicolumn{7}{@{}l@{}}{\hbox to 0pt{\parbox{110mm}{\normalsize
\par\noindent
Column 1: Observation date in 2024 (UTC); Column 2: The rest frequency of OH maser. The values 1665 and 1667 correspond to 1665.4018 and 1667.359 MHz, respectively; Column 3: Peak flux density (Jy) estimated from the Gaussian fitting (see Fig. \ref{fig:1}); Column 4: The Root Mean Squared (RMS) noise level in emission free region (Jy); Column 5: Heliocentric velocity at the peak flux density (km s$^{-1}$); Column 6: Full width at half maximum (km s$^{-1}$) estimated from the Gaussian fitting; Column 7: The ratio of the 1665 MHz line area to the 1667 MHz line area. The statistical weight ratio of 1665 to 1667 MHz emissions is 0.56.  
\tablenotetext{$*$}{In the TNRT data for March 29th, the 1665 MHz emission line was not considered a detection due to its low signal-to-noise ratio (2.5-$\sigma$) and broad linewidth (6.5 km s$^{-1}$). Thus, the line ratio cannot be determined.}
}\hss}}
\end{tabular} 
\end{center} 
\end{table*} 

\section{Results and Discussion} \label{sec:results}
We obtained statistically significant (5-$\sigma$$\sim$10-$\sigma$) detections of OH maser transitions at 1665 and 1667 MHz during the three observations (see Fig. \ref{fig:1} and Table \ref{table:2}), excluding the 1665 MHz line observed on March 29th (i.e., $<$ 3-$\sigma$). Given the broad linewidth (i.e., 6.5 km s$^{-1}$) of the 1665 MHz emission on March 29th, the marginal 1665 MHz feature
cannot be assigned to a cometary line. 
The line ratios of 1665 MHz to 1667 MHz emissions on March 22nd and 27th, 2024, are 0.69$\pm$0.10 and 0.62$\pm$0.11, respectively. Both ratios are consistent with the statistical weight ratio (i.e., 0.56) within 1.3-$\sigma$. No OH maser emission was detected at 1612 MHz. Also, strong RFI prevented us from using TNRT data to search for the 1712 MHz OH maser transition. 

As shown by Fig. \ref{fig:1} and Table \ref{table:2}, peak velocities of 1665 and 1667 MHz OH masers are in good agreement with each other in individual observations. It indicates that the 1665 and 1667 MHz emissions are unlikely due to the RFI. This is supported by the fact that observed peak velocities are in good agreement with predicted heliocentric velocities (see Table \ref{table:1}). Given the fact that the coordinate values differ by approximately 5.8 degrees between the first and second epochs (see Table \ref{table:1}), these 1665 and 1667 MHz emissions are from comet 12P/Pons-Brooks, rather than interstellar OH masers. Also, the line widths of these emissions are consistent with cometary lines, except for the 1665-MHz measurement on March 29th, 2024. Based on the TNRT observation results, we discuss the physical parameters of 12P/Pons-Brooks in the following sections.

\subsection{OH production rate Q$_{\rm{OH}}$} \label{sec:discussion-1}



\begin{table*}[htbp] 
\begin{flushleft}
\caption{OH and H$_{2}$O production rates of 12P/Pons–Brooks} 
\label{table:4}
\small 
\hspace{-1cm}
\begin{tabular}{ccccccccccc} 
\hline 
\hline 
Date & $\nu$ & \multicolumn{1}{c}{$S$}  &$f$ & $i$  	   &$l$	&$b$ &$T_{{\rm bg}}$&$\tau_{\rm{OH}}$  &$Q_{\rm{OH}}$& $Q_{\rm{H_{2}O}}$ \\
	 (mmdd)      &(MHz) &\multicolumn{1}{c}{(mJy km s$^{-1}$)} & &&(deg)&(deg)&(K)&(sec)&(10$^{29}$ sec$^{-1}$) & (10$^{29}$ sec$^{-1}$)\\ 
\hline
0322	&1665  	&607$\pm$73	&0.87 &0.43      &132&$-$34&3.3&1.01$\times$10$^{5}$ &5.35$\pm$0.64 &5.88$\pm$0.71\\

0322&1667       &880$\pm$75	&0.87&0.46  &132&$-$34&3.3&1.01$\times$10$^{5}$&3.98$\pm$0.34 &4.38$\pm$0.38\\

\multicolumn{2}{l}{Weighted mean} &$-$ &$-$&$-$&$-$&$-$&$-$&$-$ &4.28$\pm$0.30 &4.71$\pm$0.33\\
\hline

0327	&1665  	&423$\pm$59	    &0.86&0.30  &139&$-$35&3.3&0.91$\times$10$^{5}$&5.97$\pm$0.83 &6.57$\pm$0.92	\\

0327&1667       &679$\pm$67	    &0.87&0.32   &139&$-$35&3.3&0.91$\times$10$^{5}$&4.94$\pm$0.49&5.43$\pm$0.54	\\

\multicolumn{2}{l}{Weighted mean} &$-$ &$-$&$-$&$-$&$-$&$-$&$-$ &5.21$\pm$0.42 &5.73$\pm$0.46\\

\hline

0329&1667       &406$\pm$52	    &0.88&0.28 &141&$-$36&3.4	 &0.87$\times$10$^{5}$&3.36$\pm$0.43 &3.70$\pm$0.47\\

\hline 
\multicolumn{4}{@{}l@{}}{\hbox to 0pt{\parbox{160mm}{\normalsize
\par\noindent
\\
Column 1: Observation date in 2024 (UTC); Column 2: The rest frequency of OH maser. The values 1665 and 1667 correspond to 1665.4018 and 1667.359 MHz, respectively; Column 3: Integrated flux density from $-$3.5 to 3.5 km s$^{-1}$ in the cometocentric frame. Uncertainties are estimated as $\sigma$ $\times$ d$v$ $\times$ $\sqrt[]{N}$ where $\sigma$ is the rms given in Table \ref{table:2}, $N$ is the number of spectral channels and d$v$ is the channel width; Column 4: The correction factor (see the main text); Column 5: The maser inversion model of \citet{1988ApJ...331.1058S}; Columns 6-7: Approximate Galactic coordinates around the midpoint of each observation; Column 8: background temperature (see the main text); Column 9: OH lifetime (see the main text). Column 10: The production rate (molecules sec$^{-1}$) of OH radicals. For 1665 MHz line, 
the statistical weight ratio for 1665:1667 MHz (i.e., 5:9) is used to derive the production rate ; Column 11: The production rate (molecules sec$^{-1}$) of H$_{2}$O. We assume $Q_{\rm{H_{2}O}}$ = 1.1 $Q_{\rm{OH}}$ \citep{1991LPICo.765..108J}.

}\hss}}
\end{tabular} 
\end{flushleft}
\end{table*}


\begin{figure*}[tbhp] 
 \begin{center} 
     \includegraphics[scale=0.66]{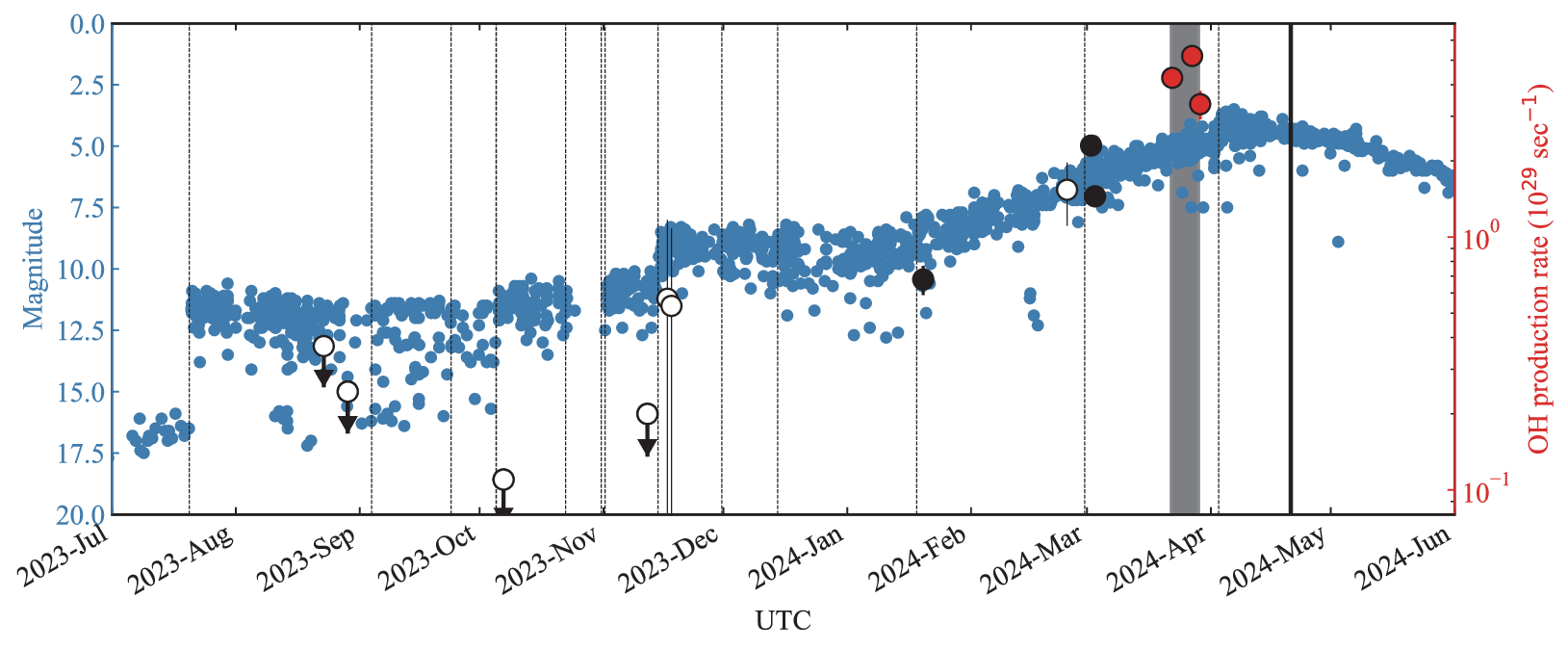} 
\end{center} 
\caption{Total visual magnitude (Y1 axis; blue circles) and OH production rate (Y2 axis; red, black and white circles) of 12P/Pons-Brooks as a function of time. The magnitude results (blue circles) are taken from COBS (Comet OBServation database), while the $Q_{\rm{OH}}$ values refer to optical (white circles; \citealp{2024ATel16498....1J,2023ATel16338....1J,2023ATel16223....1J,2023ATel16282....1J,2024MNRAS.534.1816F}) and radio results (red circles = TNRT results; black circles = \citealp{2025A&A...701A.204L}), respectively. The TNRT results obtained by averaging the 1665 and 1667 MHz results (see Table \ref{table:4}) are seen within the gray band. The dashed and bold solid lines represent the dates of optical outbursts \citep{2025arXiv250115958G} and the last perihelion, respectively. The down arrows mean upper values of OH production rates.}
\label{fig:3} 
\end{figure*}

The OH production rate ($Q_{\rm{OH}}$) can be estimated through equations (4) and (5) of \citet{1990A&A...238..382B} as 
\begin{equation}
\label{equation:1}
f\Gamma = 2.33 \times 10^{34} \frac{\Delta^{2} S}{iT_{\rm{bg}}}
\end{equation}
and
\begin{equation}
Q_{\rm{OH}} = \frac{\Gamma}{\tau_{\rm{OH}}}
\end{equation}
where $f$ is the fraction of the OH radicals at fluorescence equilibrium observed by the instrument, $\Delta$ is the topocentric distance in au, $S$ is the velocity-integrated flux density of 1667 MHz OH line in Jy km s$^{-1}$, $i$ is the maser inversion, $T_{\rm{bg}}$ is the background temperature in K, and $\tau_{\rm{OH}}$ is the OH lifetime in sec. For the 1665 MHz line, equation \ref{equation:1} needs to be modified to use a statistical weight ratio of 5:9 for 1665:1667 MHz. For an unresolved OH coma with no collisional quenching, $f$ can be assumed to be 1 \citep{2017AJ....154..249W}.


Although two OH-inversion models have been proposed by  \citet{1981A&A....99..320D} and \citet{1988ApJ...331.1058S}, we adopt the latter model in this paper. This is because the latter model is presented in tabular form. A notable difference between the two models is seen near a crossing point where the maser population goes from an inverted state to an anti-inverted state or vice versa (see \citealp{1988ApJ...331.1058S}), but this is not the case for our observations with large maser inversion values ($i \geq$ 0.28; see Table \ref{table:4}). Therefore, both models lead to similar results in our case. Regarding the background temperature $T_{\rm{bg}}$, 
we have confirmed that the comet did not pass through the Galactic plane during the three TNRT observations. Following \citet{2002A&A...393.1053C}, we estimate $T_{\rm{bg}}$ values of the TNRT observations from the Stockert continuum survey at 1420 MHz \citep{1986A&AS...63..205R}. Here we assume a 2.7 K cosmic contribution and a Galactic contribution with a spectral index of $-$2.6. Referring to the OH photodissociation time scale at 1 au ($\tau{_{\rm 1 au}}$ = 1.1 $\times$ 10$^{5}$ sec; \citealp{1990A&A...238..382B}) and a scaling law \citep{2017AJ....154..249W}, we assume an OH lifetime $\tau_{\rm{OH}}$ = $\tau{_{\rm 1 au}}$ $\times$ $r_{h}^2$ where $r_{h}$ is the heliocentric distance.



The $f$ value is estimated using the formula below:
\begin{equation}
\label{eq:4}
f = \frac{
\displaystyle \int_{r_q}^{\infty} n_d(r) \, \omega(r) \, r^2 \, dr
}{
\displaystyle \int_{0}^{\infty} n_d(r) \, r^2 \, dr
}
\end{equation}
where $n_d(r)$ is the OH number density, $\omega(r)$ the beam weighting function, $r$ the distance from the cometary nucleus, and $r_q$ the collisional quenching radius. The Haser model for daughter species \citep{1957BSRSL..43..740H, 1974A&A....34..187K} is applied to $n_d(r)$. Note that the Haser model parameters are replaced by the Haser equivalent parameters \citep{1980ApJ...237..633C}. To calculate the Haser equivalent parameters, an water expansion velocity of 1.55 km s$^{-1}$ (see the next section), an OH ejection velocity of 0.9 km s$^{-1}$ \citep{2007AA...467..729T}, and an H$_{2}$O lifetime of 4.6$\times$10$^{4}$ sec at 1 au \citep{1992Ap&SS.195....1H} are assumed. Note that the H$_{2}$O lifetime is consistent with an active Sun and scales with $r_{h}^2$. A Gaussian beam with FWHM = 1200 arcsec is applied to $\omega(r)$. The theory of OH quenching is still poorly constrained by observations \citep{2002A&A...393.1053C}. \citet{2023A&A...677A.157D} succeeded in accurately determining the OH quenching radius of comet C/2020 F3 (NEOWISE) by comparing OH production rate values obtained with different telescopes with different beam sizes (see Figure 5 of \citealp{2023A&A...677A.157D}). Referring to \citet{1990A&A...238..382B}, the $r_q$, inside which the $\Lambda$-doublet maser is quenched by collisions, is assumed to be 65,000 km at $r_h$ = 1.38 au for a production rate of 9.4$\times$10$^{28}$ sec$^{-1}$ \citep{1990A&A...230..489G} and is scaled as $r_h$($Q_{\rm{OH}}$)$^{0.5}$ \citep{1988ApJ...332..524S}. Since $r_q$ depends on $Q_{\rm{OH}}$, we determine $f$ and $Q_{\rm{OH}}$ iteratively.

Table \ref{table:4} lists OH production rate values of 12P/Pons-Brooks. 
The water production rates ($Q_{\rm{H_{2}O}}$) of the comet are also listed in the table, assuming a proportional relationship between the rates of H$_{2}$O and OH 
production, i.e., $Q_{\rm{H_{2}O}}$ = 1.1 $Q_{\rm{OH}}$ \citep{1991LPICo.765..108J}.
The OH production rates ($Q_{\rm{OH}}$) determined by TNRT observations are 4.28$\pm$0.30$\times$10$^{29}$ sec$^{-1}$, 5.21$\pm$0.42$\times$10$^{29}$ sec$^{-1}$, and 3.36$\pm$0.43$\times$10$^{29}$ sec$^{-1}$ on March 22nd, 27th and 29th, 2024, respectively. 

As a reference value, \citet{2024ATel16498....1J} reported a February 25th, 2024 OH production rate of 1.54$\pm$0.43$\times$10$^{29}$ sec$^{-1}$ for 12P/Pons-Brooks based on TRAPPIST (TRAnsiting Planets and Planetesimals Small
Telescopes) observations. \citet{2025A&A...701A.204L} also measured radio $Q_{\rm{OH}}$ values of 0.68$\pm$0.09$\times$10$^{29}$ sec$^{-1}$, 2.30$\pm$0.14$\times$10$^{29}$ sec$^{-1}$ , and 1.45$\pm$0.12$\times$10$^{29}$ sec$^{-1}$  on January 20th, March 2nd, and March 3rd, 2024, respectively. They argued that the 37$\%$ decrease in $Q_{\rm{OH}}$ value between March 2nd and 3rd could be related to the optical outburst on February 29th, 2024. Although no optical outburst was observed between February 29th and April 2nd, 2024, the TNRT results confirm a 36$\pm$10$\%$  decrease in $Q_{\rm{OH}}$ value between March 27th and 29th, 2024. Figure \ref{fig:3} shows the total visual magnitude and OH production rate of 12P/Pons-Brooks as a function of time. The total visual magnitude progressively decreases as 12P/Pons-Brooks approaches the last perihelion on April 21, 2024 although there are discontinuities in the time variation of the total visual magnitude due to optical outbursts (see Fig. \ref{fig:3}). The well documented outbursts are highlighted by dashed lines in Fig. \ref{fig:3}. Although the number of $Q_{\rm{OH}}$ values is limited, $Q_{\rm{OH}}$ progressively increases as 12P/Pons-Brooks approaches the last perihelion. 


\begin{figure*}[tbhp] 
 \begin{center} 
     \includegraphics[scale=1.2]{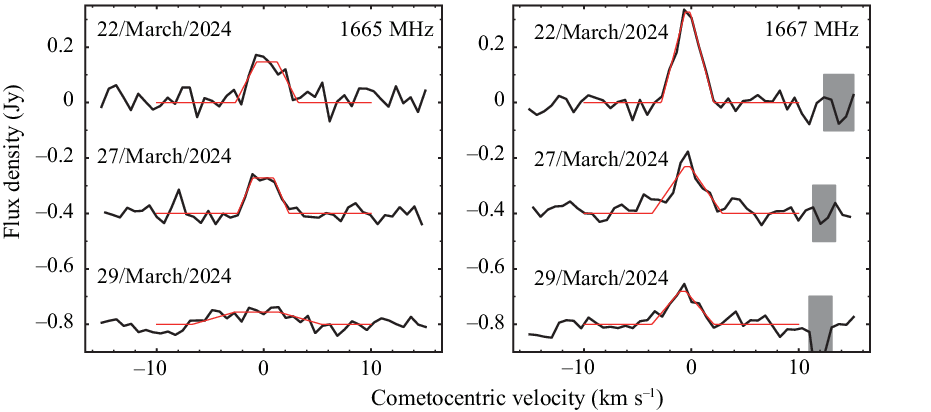} 
\end{center} 
\caption{Similar to Figure \ref{fig:1}, but shown in the cometocentric frame. Red shows symmetric trapezoid fits to the data, instead of Gaussian fits (see the text for details). Note that no 1665 MHz emissions were detected on March 29th, 2024.}
\label{fig:2} 
\end{figure*}  


\begin{table*}[htbp] 
\begin{center}
\caption{Symmetric trapezoid fits to the 18-cm OH line profiles of 12P/Pons–Brooks} 
\label{table:3}
\small 
\begin{tabular}{ccrcccccc} 
\hline 
\hline 
Date & $\nu$ & \multicolumn{1}{c}{$V_{0}$} &\bf{$A$} &$w_{\rm{rise}}$ &$w_{\rm{flat}}$ &$\sigma$ &2($V_{p}$+$V_{d}$)  & $V_{p}$ 	    	  \\
	 (mmdd)      &(MHz) &\multicolumn{1}{c}{(km s$^{-1}$)}&(mJy) &(km s$^{-1}$) &(km s$^{-1}$) &(mJy) &(km s$^{-1}$) &(km s$^{-1}$)\\ 
\hline
0322	&1665  	&0.28$^{+0.22}_{-0.28}$&147$^{+26}_{-22}$	&1.97$\pm$1.29 &1.91$^{+1.37}_{-1.31}$&38$^{+6}_{-5}$&5.57$^{+2.01}_{-1.05}$ &1.89$^{+1.00}_{-0.53}$	  \\

0322&1667&$-$0.35$^{+0.07}_{-0.06}$	&326$^{+27}_{-23}$ &2.21$^{+0.21}_{-0.26}$&0.47$^{+0.36}_{-0.32}$&24$^{+4}_{-3}$&4.89$^{+0.27}_{-0.30}$ &1.54$^{+0.14}_{-0.15}$\\

\multicolumn{2}{l}{Weighted mean}& \multicolumn{1}{c}{$-$} &$-$&  $-$ &$-$&$-$&$-$&1.55$\pm$0.14 \\
\multicolumn{2}{l}{Heuristic law\tablenotemark{$*$}}& \multicolumn{1}{c}{$-$} &$-$&  $-$ &$-$&$-$&$-$&1.58$^{+0.09}_{-0.08}$ \\
\hline
0327	&1665  	&$-$0.06$\pm$0.15	&128$^{+17}_{-15}$ &1.43$^{+0.77}_{-0.58}$ &1.94$^{+0.67}_{-0.86}$&26$^{+4}_{-3}$&4.74$^{+0.82}_{-0.63}$&1.47$^{+0.41}_{-0.32}$	\\

0327&1667   &$-$0.38$^{+0.20}_{-0.21}$	&169$^{+31}_{-32}$ &3.02$^{+1.68}_{-0.76}$ &0.50$^{+0.67}_{-0.36}$&32$^{+5}_{-4}$&6.61$^{+3.39}_{-1.46}$&2.41$^{+1.70}_{-0.73}$	\\

\multicolumn{2}{l}{Weighted mean}& \multicolumn{1}{c}{$-$} &$-$ & $-$ &$-$ &$-$ &$-$&1.55$\pm$0.35 \\
\multicolumn{2}{l}{Heuristic law\tablenotemark{$*$}}& \multicolumn{1}{c}{$-$} &$-$&  $-$ &$-$&$-$&$-$&1.69$^{+0.10}_{-0.09}$ \\
\hline

0329&1667&$-$0.79$^{+0.16}_{-0.13}$&119$\pm$17	&2.61$^{+0.49}_{-0.60}$ &0.56$^{+0.77}_{-0.40}$&21$^{+4}_{-3}$&5.85$^{+0.94}_{-0.90}$ &2.02$^{+0.47}_{-0.45}$	  \\

\multicolumn{2}{l}{Heuristic law\tablenotemark{$*$}}& \multicolumn{1}{c}{$-$} &$-$&  $-$ &$-$&$-$&$-$&1.56$^{+0.10}_{-0.09}$ \\
\hline 
\multicolumn{9}{@{}l@{}}{\hbox to 0pt{\parbox{143mm}{\normalsize
\par\noindent
\\
Column 1: Observation date in 2024 (UTC); Column 2: The rest frequency of OH maser. The values 1665 and 1667 correspond to 1665.4018 and 1667.359 MHz, respectively; Column 3: The center velocity of the symmetric trapezoid; Column 4: The amplitude (height) of the symmetric trapezoid; Column 5: Symmetric trapezoid rise/fall width; Column 6: Width of the flat part of the symmetric trapezoid; Column 7: The noise level required to achieve a reduced chi-squared value of approximately unity after model fitting (see the text for details); Column 8: The large base of the symmetric trapezoid fit is assumed to be 2($V_{p}$+$V_{d}$) where $V_{p}$ and $V_{d}$ are water expansion velocity and OH ejection velocity, respectively; Column 9: Water expansion velocity. 
\tablenotetext{$*$}{It refers to \citet{2007AA...467..729T}.} %
}\hss}}
\end{tabular} 
\end{center} 
\end{table*}



\begin{table}[htbp]
\centering

\caption{\centering $V_p$ 1-$\sigma$ uncertainties (km s$^{-1}$) for various S/N and $\Delta V$ combinations\tablenotemark{$*$}}
\label{tab:vp_error_summary}
\setlength{\tabcolsep}{6pt}      
\begin{tabular}{|c|c|c|c|c|c|c|}
\hline
\shortstack{$\Delta V$ \\ (km s$^{-1}$)} & S/N = 2  & 4 & 5 & 6 & 10 & 14 \\
\hline
0.1 & 0.39 & 0.17 & 0.13 & 0.11 & 0.07 & 0.05 \\
\hline
0.4 &2.12  &0.40  & 0.31 & 0.25 & 0.14 & 0.09 \\
\hline
0.7 &\multicolumn{1}{>{\columncolor{gray!20}}c|}{3.48} &\multicolumn{1}{>{\columncolor{gray!20}}c|}{0.66} &\multicolumn{1}{>{\columncolor{gray!20}}c|}{0.44} &\multicolumn{1}{>{\columncolor{gray!20}}c|}{0.34} & 0.18 &\multicolumn{1}{>{\columncolor{gray!20}}c|}{0.13} \\
\hline
1.4 &5.20  &2.01  & 1.14 &0.73  &0.29  & 0.19 \\
\hline
\multicolumn{7}{@{}l@{}}{\hbox to 0pt{\parbox{85mm}{\normalsize
\par\noindent
\tablenotetext{$*$}{S/N is the signal-to-noise ratio of the OH maser emission and $\Delta V$ is the velocity resolution. We apply trapezoidal model parameters $V_{0}$ = 0.0 km s$^{-1}$, $w_{\rm{rise}}$ = 2.0 km s$^{-1}$, $w_{\rm{flat}}$ = 0.5 km s$^{-1}$, and $A$ = 300 mJy for MCMC simulations. The shaded areas correspond to the ($\Delta V$, S/N) values of the TNRT results (i.e., Table \ref{table:3}).} %
}\hss}}
\end{tabular}
\end{table}

\subsection{H$_{2}$O expansion velocity} \label{sec:discussion-2}

The expansion velocity of OH-parent (i.e., H$_{2}$O) in the cometary coma can be retrieved from the 18-cm OH line profile \citep{1990A&A...238..382B}. By fitting a symmetric trapezoid to an OH spectrum, the large base of the trapezoid is expected to be close to 2($V_{p}$+$V_{d}$) where $V_{p}$ and $V_{d}$ are the water expansion velocity and the OH ejection velocity, respectively. As in \citet{2007AA...467..729T}, we assume $V_{d}$ = 0.9 km s$^{-1}$, which is close to the theoretical value of 1.05 km s$^{-1}$ \citep{1989A&A...213..459C}. To fit a symmetric trapezoid to each TNRT result, 
we use the python package \texttt{emcee} which is an MIT licensed pure-Python implementation of Goodman $\&$ Weare's Affine Invariant Markov Chain Monte Carlo (MCMC) Ensemble sampler (\citealp{2013PASP..125..306F}). Further details on the symmetric trapezoidal model are given in the Appendix \ref{appendix:A}.

The results of a symmetric trapezoid fit to 18-cm OH spectra of 12P/Pons-Brooks are summarized in Table \ref{table:3} and Figure \ref{fig:2}. The determined water expansion velocities exhibit large uncertainties (i.e., $>$ 0.3 km s$^{-1}$), except for $V_{p}$ = 1.54$^{+0.14}_{-0.15}$ km s$^{-1}$, which is based on a 1667-MHz OH spectrum taken on March 22nd, 2024. To understand why the errors in the water expansion velocity in Table \ref{table:3} vary widely, we perform a simulation for various S/N and $\Delta V$ combinations in Table \ref{tab:vp_error_summary} where S/N is the signal-to-noise ratio of the OH maser emission and $\Delta V$ is the velocity resolution. The MCMC fitting is repeated on the simulated data to estimate the 1-$\sigma$ uncertainty in the water expansion velocity ($V_p$) for each combination of (S/N, $\Delta V$). The number of simulations for each combination is 100 or more, taking into consideration the convergence of the results. Table \ref{tab:vp_error_summary} indicates that the difference in the uncertainties of $V_{p}$ in Table \ref{table:3} is due to the difference in S/N values. Simulation results also suggest that applying channel averaging to TNRT data cannot improve the uncertainty in $V_p$.

\citet{2007AA...467..729T} provided a heuristic formulation of the water expansion velocity ($V_{p}$) as a function of heliocentric distance $r_{h}$ and OH production rate ($Q_{\rm{OH}}$) as
\begin{equation}
\label{eq:3}
V_{p} = 1.11 \pm 0.02 \times r_{h}^{-0.45\pm0.05} \times Q_{\rm{OH}}^{0.23\pm0.01} 
\end{equation}
for 0.3 $<$ $r_{h}$ (au) $<$ 1.0 and 10$^{28}$ $<$ $Q_{\rm{OH}}$ (sec$^{-1}$) $<$ 10$^{30}$. If we apply the TNRT observation results of March 22nd, 2024, $r_{h}$ = 0.959 au and $Q_{\rm{OH}}$ = 4.28$\pm$0.30$\times$10$^{29}$ sec$^{-1}$, to equation \ref{eq:3}, the heuristic expansion velocity is $V_{p}$ = 1.58$^{+0.09}_{-0.08}$ km s$^{-1}$. Since $V_{p}$ is most accurately constrained using the TNRT spectra for March 22nd, 2024, we focus on those spectra. The expansion velocity determined from the TNRT data on the same day (see Table \ref{table:3}) is 1.55$\pm$0.14 km s$^{-1}$, which is consistent with the heuristic result within errors. This 
indicates that the water expansion velocity and the OH production rate obtained from the 
TNRT data on March 22nd are in good agreement with values for other comets at similar 
heliocentric distances.

\section{Summary} \label{sec:summary}
We observed the periodic comet 12P/Pons-Brooks with 40-m TNRT on March 22nd, 27th and 29th, 2024, and succeeded in detecting the 1665 and 1667 MHz OH maser emissions in the comet (Figures \ref{fig:0} and \ref{fig:1}; Tables \ref{table:1} and \ref{table:2}). Based on the observational results, the OH production rates ($Q_{\rm{OH}}$) of 12P/Pons-Brooks were determined as 4.28$\pm$0.30 $\times$ 10$^{29}$ sec$^{-1}$, 5.21$\pm$0.42 $\times$ 10$^{29}$ sec$^{-1}$, and 3.36$\pm$0.43 $\times$ 10$^{29}$ sec$^{-1}$ for March 22nd, 27th and 29th, respectively (Figure \ref{fig:3}; Table \ref{table:4}). The TNRT results confirm a 36$\pm$10$\%$ decrease in $Q_{\rm{OH}}$ value between March 27th and 29th, 2024.
Combining these with previous estimates (\citealp{2024ATel16498....1J,2023ATel16338....1J,2023ATel16223....1J,2023ATel16282....1J, 2024MNRAS.534.1816F, 2025A&A...701A.204L}), we find that the OH production rate of 12P/Pons-Brooks gradually increases on longer timescale as the comet approaches the last perihelion on April 21st, 2024. Combining radio observations results of 12P/Pons-Brooks may reveal the existence of radio outbursts since optical outbursts have been reported for the comet \citep{2025arXiv250115958G}. 

We also determined the water expansion velocities of 12P/Pons-Brooks based on the spectral line shapes of the 1665 and 1667 MHz OH data (Figure \ref{fig:2}; Table \ref{table:3}). Although the uncertainties of the expansion velocities ($V_{p}$) were large, due to differences in the signal-to-noise ratios of the OH maser emissions (see Table \ref{tab:vp_error_summary}), the most accurate result, $V_{p}$ = 1.55$\pm$0.14 km s$^{-1}$, is consistent with a heuristic law's result of 1.58$^{+0.09}_{-0.08}$ km s$^{-1}$ \citep{2007AA...467..729T} within 0.2-$\sigma$. 

Even though the measurements of OH production rates for comet 12P/Pons-Brooks were made over a very short period of time using the 40-m TNRT, these were the first measurements using the telescope. We plan to make more intensive observations with the 40-m TNRT of several comets near perihelion each year in order to generate the TNRT cometary database (TNRT CDB). Since the H$_{2}$O production rates converted from the OH production rates are key data for many cometary analyses, our database will be used for various purposes in cometary analyses.



\section*{Acknowledgments}



We thank several reviewers for their comments, which improved the quality of the manuscript. The 40-m Thai National Radio Telescope is based on the national flagship project for human capacity building and technology development through radio astronomy and geodesy in Thailand, which is supported by Ministry of Higher Education, Science, Research and Innovation (MHESI) for the telescope construction and the development, upgrade and installation of receivers. The construction of the 40-m Thai National Radio Telescope and the development of its receivers were financially supported by the fiscal year 2019 expenditure budget allocated by the Bureau of the Budget, Thailand, through NARIT.

Authors specially thank Michael Kramer, G. W., E. B., Christoph Kasemann, N. A. E., J. W., T. W., Jan Behrend, and Rafael Castenholz of Max-Planck-Institut f\"{u}r Radioastronomie for the development of the L- and K-band receivers of the 40-m TNRT and the Universal Software Backend (USB) system using the Effelsberg Direct Digitization (EDD), and USB upgrade, which is led by M.K., G.W., and E.B.. 

We appreciate Jos{\'e} Antonio L{\'o}pez-P{\'e}rez in Yebes Observatory team, IGN, for the development of Telescope Control Software based on ALMA Common Software, which is led by P. d. V.. We would also like to thank Jos{\'e} Antonio L{\'o}pez-Fern{\'a}ndez, former director of the Yebes Observatory, for his support in the early stages of TNRT.

The authors would like to express their sincere gratitude to all staff who have retired from Thai National Radio astronomy Observatory since its establishment. 

 We also thank the members of the International Technical Advisory Committee and the International Scientific Advisory Committee (ITAC/ISAC) for their supportive and fruitful suggestions and advice: (ITAC members)
    H. K. (Chair); B. H. K.; Do-Young Byun; Francisco P. Colomer; Agnieszka S{\l}owikowska; Michael Garrett; Yashwant Gupta; Mareki Honma; Jinling Li; Young Chol Minh; Zhiqiang Shen; Tasso Tzioumis; P.d.V.; G. W.; Kee-Tae Kim. (ISAC members) Michael Bode (Chair); B.H.K.; H.K.; M.K..
    
  This work is supported by the Fundamental Fund of Thailand Science Research and Innovation (TSRI) through the National Astronomical Research Institute of Thailand (Public Organization) (FFB680072/0269).


%

\vspace{5mm}

\facilities{COBS}




\software{Astropy \citep{2013A&A...558A..33A,2018AJ....156..123A}; emcee \citep{2013PASP..125..306F}
          }



\appendix

\section{Symmetric trapezoid model}
\label{appendix:A}
The symmetric trapezoid model used in the text is shown as
\begin{equation}
\label{eq:2b}
f(V) = \left\{
\begin{array}{ll}
0, & V < V_{0} - \frac{w_{\rm flat}}{2} - w_{\rm rise} \\[6pt]
A  \left(\frac{V - \left(V_{0} - \frac{w_{\rm flat}}{2} - w_{\rm rise}\right)}{w_{\rm rise}}\right), & V_{0} - \frac{w_{\rm flat}}{2} - w_{\rm rise} \leq V < V_{0} - \frac{w_{\rm flat}}{2} \\[6pt]
A, & V_{0} - \frac{w_{\rm flat}}{2} \leq x < V_{0} + \frac{w_{\rm flat}}{2} \\[6pt]
A  \left(1 - \frac{V - \left(V_{0} + \frac{w_{\rm flat}}{2}\right)}{w_{\rm rise}}\right), & V_{0} + \frac{w_{\rm flat}}{2} \leq V < V_{0} + \frac{w_{\rm flat}}{2} + w_{\rm rise} \\[6pt]
0, & V \geq V_{0} + \frac{w_{\rm flat}}{2} + w_{\rm rise}
\end{array}
\right.
\end{equation}
where $A$ is the height, $V_{0}$ is the center position, $w_{\rm flat}$ and $w_{\rm rise}$ are the widths of the flat and rise/fall parts respectively. In equation \ref{eq:2b}, if $w_{\rm rise}$ approaches 0, the trapezoidal function is switched to a rectangular function to make the fitting numerically more stable. We make the 1-$\sigma$ value of the Gaussian likelihood a variable so that the reduced chi-squared value for each MCMC fit approaches unity.



\bibliography{sample631}{}
\bibliographystyle{aasjournal}



\end{document}